\documentstyle[aps,prb,epsf,preprint]{revtex}
\tighten
\preprint{IASSNS-HEP 00/61}
\title{\bf Parity (and time-reversal) anomaly in a semiconductor}
\author{Oleg Tchernyshyov\cite{e-mail}}
\address{School of Natural Sciences, Institute for Advanced Study, 
Princeton, New Jersey 08540\\
Argonne National Laboratory, 
9700 South Cass Avenue, Argonne, Illinois 60439}
\date{July 7, 2000}

\begin{document}
\maketitle

\begin{abstract}
The physics of a parity anomaly, potentially observable in a
narrow-gap semiconductor, is revisited.  Fradkin, Dagotto, and
Boyanovsky have suggested that a Hall current of anomalous parity can
be induced by a Peierls distortion on a domain wall.  I argue that a
perturbation inducing the parity anomaly must break the time reversal
symmetry, which rules out the Peierls distortion as a potential cause.
I list all possible perturbations that can generate the anomaly.
\end{abstract}

\section{Introduction}

Condensed-matter and particle physics are cross-fertilizing fields.
For instance, exclusion of magnetic flux from a superconductor was
interpreted by Anderson as generation of mass for a gauge
boson.\cite{Anderson} Popularized in particle physics by
Higgs,\cite{Higgs} this effect plays a central role in the electroweak
theory.  In the other direction, statistical (Chern-Simons) gauge
fields, a field-theorist's toy, facilitate better understanding of the
quantum Hall effect.\cite{Wilczek} A closely related phenomenon,
parity anomaly in $(2+1)$-dimensional electrodynamics, still awaits
its experimental discovery in condensed matter.

In a nutshell, the effect is rather simple. Two-component Dirac
fermions with a mass $m$ react in a peculiar way to an external
magnetic field $B$.  The symmetry between occupied (Fermi-sea) and
empty fermion states is violated and a nonzero fermion density appears
in the vacuum state.  The density of fermions---each carrying charge 
$\pm e/2$---is such that the average flux per particle is $hc/e$:
\begin{equation}
\rho = \frac{Q}{A} = \frac{m}{|m|}\frac{e^2 B}{2hc}.
\label{CS-0}
\end{equation}
By relativistic invariance, application of an electric field in the
plane induces a Hall current with quantized conductivity: 
\begin{equation}
J_i = (\sigma/c) \epsilon_{ij}E_j, 
\hskip 5mm 
\sigma = -\frac{m}{|m|}\frac{e^2}{2h}.
\label{CS-1}
\end{equation}
Both the induction of charge by magnetic field (\ref{CS-0}) and the
appearance of a Chern-Simons current (\ref{CS-1}) violate the
symmetries of parity and time reversal, hence the name ``parity
anomaly''.  In certain materials, electrons behave as Dirac fermions,
albeit with a small ``speed of light'', so it seems natural to look
for the anomaly in condensed matter.

On a deeper level, there is a subtle problem, known as fermion
doubling, which often prevents the anomaly.  For example, electron
states in a sheet of graphite are well described at low energies as
2-component massless Dirac fermions in 2+1 dimensions.  Because the
unit cell contains 2 lattice sites, there are 2 fermion species living
at inequivalent Fermi points in the Brillouin zone.  Semenoff observed
that a symmetry breaking mass term $m$ will be induced if the
inequivalent sites are populated by different atoms.  Unfortunately,
the 2 species have mass terms of opposite signs.\cite{Semenoff}  The
total anomalous current (\ref{CS-1}) vanishes.

Later, Fradkin, Dagotto and Boyanovsky\cite{FDB} (FDB) suggested a
possible way to circumvent the problem of fermion doubling. Lead
chalcogenides PbTe, PbSe, and PbS are semiconductors with a narrow gap
($0.15$--$0.3$ eV) between conduction and valence
bands.\cite{Nimtz-Schlicht} Low-energy fermion quasiparticles are
concentrated around 4 inequivalent L points in the Brillouin zone,
$\pm(\pi/a)(1,1,1)$ and three others related by cubic symmetry.  The
band structure near each L point is such that the quasiparticles
resemble 4-component massive Dirac fermions (in 3+1 dimensions).  4
components result from 2 inequivalent sites in a unit cell and 2
possible projections of a spin $1/2$.  FDB pointed out that a stack
fault in a crystal creates a domain wall, on which one finds 2+1
dimensional massless Dirac fermions.  In this setting, certain
perturbations can induce a mass term of {\em the same} sign in all 4
fermion species.  The anomalous current (\ref{CS-1}) does not have to
vanish.

In a more detailed paper,\cite{BDF} FDB suggested that a
symmetry-breaking mass term $m$ can be induced by a Peierls
distortion, which is present in some lead
chalcogenides.\cite{Nimtz-Schlicht} That does not sound right: a
Peierls distortion violates parity but leaves time reversal intact,
therefore it cannot possibly induce a Chern-Simons current
(\ref{CS-1}).  Haldane has argued that this incarnation of parity
anomaly is caused by an unphysical lattice Hamiltonian used by FDB,
which is (or seems to be) odd under time reversal.\cite{Haldane} Upon
further reflection, this argument does not work: if a continuum limit
derives from a model with lattice potential and spin-orbit
interaction, it cannot violate the time-reversal symmetry.  Then why
does the FDB Hamiltonian couple spin and linear momentum via the term
$\vec\sigma\!\cdot\!{\bf k}$ --- that seems to break $T$?  The answer:
the $\sigma$ matrices here are {\em not} the electron spin variables.
Symmetry under time reversal need not be broken.  The existence of
the anomalous Chern-Simons current is then questionable.  

Being field theorists, FDB didn't bother to derive their Hamiltonian
(nor its continuum version) from any physical model of electrons in a
lattice potential and with the spin-orbit interaction.  Their
phenomenological Hamiltonian was merely tailored to correctly
reproduce the energy spectrum of the conduction and valence bands near
the Fermi level.  Therefore the $\sigma$ matrices in it have nothing
to do with the actual electron spin.  But without knowing which
physical variables the $\sigma$ matrices represent, one cannot learn
how this or that physical perturbation (e.g., a Peierls distortion)
couples to the fermion zero modes.  Then it is impossible to determine
correctly the sign of charge---or the direction of Hall
current---induced by a symmetry-breaking perturbation.

To clear up the matter, I have derived a correct continuum
approximation for low-lying electron states both by appealing to
symmetry arguments (the easy way) and by using the standard ${\bf
k\!\cdot\!p}$ method combined with a tight-binding
approximation\cite{Slater-Koster,Mitchell-Wallis,VPS} (see Appendix).
I have studied the behavior of massless fermions bound to a domain
wall and enumerated {\em all} symmetry-breaking perturbations that
raise or lower the zero modes in magnetic field, thus inducing surface
charge (\ref{CS-0}).  The results of this study are unambiguous.
Already from Eq. \ref{CS-0} it is clear that $m$ should be odd under
both parity and time reversal (a pseudoscalar).  A Peierls distortion
therefore will not work because $m={\bf q\!\cdot\!\hat{n}}$ is a
genuine scalar (${\bf q}$ is the vector of atomic displacement and
$\hat{\bf n}$ is the normal to the domain wall).  On the other hand,
certain kinds of magnetic order may induce an anomalous Hall current.

\section{Low-energy Hamiltonian}

The FDB Hamiltonian (in the continuum limit) can be derived from a
tight-binding model of noninteracting electrons that includes electron
kinetic energy, periodic lattice potential, and spin-orbit
interaction:\cite{Mitchell-Wallis}
\begin{equation}
H = \frac{p^2}{2m} + V({\bf r}) 
+ \frac{[\vec\sigma\!\times\!\nabla V({\bf r})]\cdot {\bf p}}{4m^2c^2}.
\end{equation}
This Hamiltonian is invariant under parity and time reversal.  

The triplet of Pauli matrices $\vec\sigma$ denotes the spin operators.
Introduce operators of isospin $\tau_1, \tau_2, \tau_3$ to describe
the two sublattices in the rocksalt structure.  Eigenvalues
$\tau_3=\pm1$ correspond to lead and chalcogen sites.  Near each of
the four L points, lowest-energy states have a 4-component wavefunction
(2 spin components $\times$ 2 sublattices) and resemble Dirac
fermions.  By symmetry arguments alone, one can anticipate the correct
form of the Hamiltonian in the continuum limit.

Define the operations of parity and time reversal as
\begin{equation}
\begin{array}{cc}
P \psi({\bf r}) = -\psi(-{\bf r}), &
T \psi({\bf r}) = \sigma_y \psi^*({\bf r}).
\end{array}
\label{PT}
\end{equation}
The minus sign accounts for odd parity of $p$ orbitals.  Near 
an L point in the Brillouin zone, e.g., ${\bf p} = (\pi/a)(1,1,1)$,
the one-particle Hamiltonian of a Dirac fermion is limited to the
following $P$ and $T$-invariant terms:
\begin{equation}
H = -iv \tau_1 (\hat{\bf p} \cdot \nabla)
-iv\lambda\tau_2 ([\vec\sigma \times \hat{\bf p}] \cdot \nabla)
+\tau_3 Mv^2.
\label{Dirac3+1}
\end{equation}
Here $\hat{\bf p} = {\bf p}/p$ is odd under both parity and time
reversal (directions $\hat{\bf p}$ and $-\hat{\bf p}$ describe the
same L point); $\lambda$ is the relative strength of the spin-orbit
interaction, and $v$ is a fermion velocity.  For simplicity, I set
$\lambda=1$, which gives a spherically symmetric energy spectrum near
the L points.  The same form of $H$ is obtained from a tight-binding
model, see Appendix \ref{appendix}.  The low-energy Hamiltonian thus
has a Dirac form $H=-i\vec\alpha\!\cdot\!\nabla + \beta M$.  The
standard Dirac matrices have the following representation:
\begin{equation}
\begin{array}{ll}
\vec{\alpha} 
= \tau_1 \hat{\bf p} + \tau_2 [\vec\sigma \times \hat{\bf p}], &
\beta = \gamma^0 = \tau_3, \\
-i\vec{\gamma} = -i\beta\vec{\alpha} 
= \tau_2 \hat{\bf p} - \tau_1 [\vec\sigma \times \hat{\bf p}], &
\gamma^5 = -\tau_1(\vec\sigma\!\cdot\!\hat{\bf p}), \\
\vec\Sigma = -\gamma^5\vec\alpha 
= (\vec\sigma\!\cdot\!\hat{\bf p})\hat{\bf p}
+ \tau_3\, \hat{\bf p}\!\times\![\vec\sigma\!\times\!\hat{\bf p}], &
i\gamma^0\gamma^5 =  \tau_2(\vec\sigma\!\cdot\!\hat{\bf p})
\end{array}
\end{equation}

\section{Massless fermions on a domain wall}

Consider a stack fault (Pb $\leftrightarrow$ Te) in a plane
perpendicular to a unit vector $\hat{\bf n}$.  Then $M$ is a function
of $x_{||} = {\bf r}\!\cdot\!\hat{\bf n}$.  More specifically,
the normal to the domain wall $\hat{\bf n}$ points in the direction of
increasing $M$:  
\begin{equation}
\begin{array}{cc}
M(x_{||}) \to \pm |M_0| &
\mbox{for } x_{||} \equiv {\bf r\!\cdot\!\hat{n}} \to \pm\infty,
\end{array}
\label{n-direction}
\end{equation}
Energy eigenstates bound to the domain wall can be written in the form
$\psi({\bf r}) = u(x_{||})\psi({\bf x}_\perp)$, where 
${\bf x}_\perp = \hat{\bf n}\!\times\![{\bf r}\!\times\!\hat{\bf n}]$
are coordinates within the plane.  The scalar $u(x_{||})$ and bispinor
$\psi({\bf x}_\perp)$ satisfy the following equations:\cite{FDB}
\begin{eqnarray}
u'(x_{||}) = -M(x_{||})\,u(x_{||}), 
	\label{eqn-x-par}\\
(-i\vec\gamma\!\cdot\!\hat{\bf n})\, \psi({\bf x}_{\perp}) 
	= \psi({\bf x}_\perp), 
	\label{gamma-n-eigen}\\
-iv\, \vec\alpha\! \cdot\! \nabla_{\!\perp}\psi({\bf x}_\perp) 
	= E\, \psi({\bf x}_\perp)
	\label{eqn-x-perp} 
\end{eqnarray}
The first two equations (\ref{eqn-x-par}--\ref{gamma-n-eigen})
indicate that these states are fermion zero modes in the direction
perpendicular to the wall.\cite{Jackiw-Rebbi} The last two
(\ref{gamma-n-eigen}-\ref{eqn-x-perp}) describe 2-component massless
fermions in 2+1 dimensions with the Hamiltonian
\begin{equation}
H = v\, \vec\alpha\! \cdot\! (-i\nabla_{\!\perp} - \frac{e}{c}{\bf A}_\perp).
\end{equation}

Application of magnetic field ${\bf B} = B\hat{\bf n}$ perpendicular
to the domain wall results in a spectrum with Landau levels:\cite{Zeeman}
\begin{equation}
H^2 = v^2\left(-i\nabla_{\!\perp} - \frac{e}{c}{\bf A}_\perp\right)^2 
- \frac{eBv^2}{c}\, \vec\Sigma\!\cdot\!\hat{\bf n}.
\end{equation}
The energy spectrum is symmetric with respect to charge conjugation:
\begin{equation}
E = 0,\, \pm v(2|eB|/c)^{1/2},\, \pm v(4|eB|/c)^{1/2}\ldots 
\end{equation}
Zero modes ($E=0$) are eigenstates of the ``spin'' component
$\vec\Sigma\!\cdot\!\hat{\bf n}$ with the eigenvalue $\mbox{sgn}(eB)$.
All Landau levels have orbital degeneracy $|eB|/2\pi c$ per unit area.

In a half-filled system (no dopants), the zero modes are exactly at
the Fermi level and thus have occupation numbers $1/2$.  The energy
spectrum is particle-hole symmetric and the domain wall is not
charged.  An arbitrarily small perturbation can shift the zero modes
above or below the chemical potential (assuming it stays pinned at 0).
With the particle-hole symmetry broken, the domain wall gets charged:
each zero mode contributes charge $+e/2$ (empty) or $-e/2$ (filled) to
the domain wall ($e<0$).\cite{Jackiw-Rebbi}

Although a condensed-matter system lacks the true relativistic
invariance and quantization of Hall conductivity (\ref{CS-1}) does not
follow automatically from Eq.~\ref{CS-0}, there is a thermodynamic
identity that establishes this relation:\cite{Streda}
\begin{equation}
\sigma = -\left.\frac{\partial\rho}{\partial B}\right|_{\mu=\rm const}
= -\frac{m}{|m|}\frac{e^2}{2h},
\label{CS-1CM}
\end{equation}
with the derivative taken at a constant chemical potential.  This
result is valid in the absence of low-energy excitations.

\section{Symmetry-breaking perturbations}

Imagine now that a uniform perturbation $V$ (such as a Peierls
distortion or the actual Zeeman term) is applied to fermions on the
domain wall.  In the presence of the Landau gap, the shift of the
zero modes can be computed to first order in $V$ using the standard
perturbation theory.  Because there is no spin or isospin degeneracy
and $V$ is uniform, $\Delta E = \int\psi^\dagger V \psi\ d^3r$. 

Recall that $\psi$ is an eigenstate of 
3 commuting matrices:
\begin{eqnarray}
(\vec\Sigma\!\cdot\!\hat{\bf n})\, \psi &=& \mbox{sgn}(eB)\, \psi, 
\label{Sigma-n}\\
(-i\vec\gamma\!\cdot\!\hat{\bf n})\, \psi  &=& \psi, 
\label{gamma-n}\\
(i\gamma^0\gamma^5)\, \psi &=& \mbox{sgn}(eB)\, \psi. 
\label{gamma-05}
\end{eqnarray}
(The third matrix $\gamma^0\gamma^5$ is simply the product of the
first two.)  To compute $\psi^\dagger V \psi$, one can write $V$ as a
superposition of 15 traceless Hermitian $4\!\times\!4$ matrices 
\[\{H_i\} =
\{\gamma^0,\  \gamma^5,\  i\gamma^0\gamma^5,\  \vec\alpha,\  
-i\vec\gamma,\  \vec\Sigma,\  \gamma^0\vec\Sigma\}:
\]
\begin{equation}
\psi^\dagger V \psi 
= \sum_{i=1}^{15} \frac{1}{4}\mbox{Tr}(V H_i)\, \psi^\dagger\! H_i \psi 
\label{15terms}
\end{equation}
Because operators $H_i$ either commute or anticommute with one
another, only 3 of them (\ref{Sigma-n}--\ref{gamma-05}) need to be
included in the sum: if some $H_j$ anticommutes with one of them, its
expectation value in the state $\psi$ vanishes.

Thus operators (\ref{Sigma-n}--\ref{gamma-05}) exhaust all the handles
through which external perturbations can tickle the zero modes and
possibly induce charge on a domain wall.  Such a perturbation will
contain ${\bf F}_\Sigma\!\cdot\!\vec\Sigma$, $-i{\bf
F}_\gamma\!\cdot\vec\gamma$, or $iF_5 \gamma^0\gamma^5$.  The
symmetry breaking term ${\bf F}_\Sigma$ is a pseudovector, ${\bf
F}_\gamma$ is a vector, and $F_5$ is a pseudoscalar.  The area density
of charge induced by them on a domain wall (for a single fermion
species) is
\begin{equation}
- \mbox{sgn}({\bf F}_\Sigma\!\cdot\!\hat{\bf n})
\,\frac{e^2({\bf B}\!\cdot\!\hat{\bf n})}{4\pi c\hbar}, 
\hskip 3mm
- \mbox{sgn}({\bf F}_\gamma\!\cdot\!\hat{\bf n})\,\frac{e|eB|}{4\pi c\hbar},
\hskip 3mm
- \mbox{sgn}(F_5)\, \frac{e^2({\bf B}\!\cdot\!\hat{\bf n})}{4\pi c\hbar}. 
\end{equation}

It is now evident that a vector perturbation ${\bf F}_\gamma$, such as
a Peierls distortion, cannot induce a Chern-Simons current: the sign
of induced charge is not sensitive to the direction of magnetic field.
By the thermodynamic identity (\ref{CS-1CM}), there will be a normal
Hall effect with current reversing the direction when magnetic field
does.

One obviously needs a pseudoscalar perturbation (odd under both $P$
and $T$) in order to relate a scalar (charge) to the pseudoscalar
${\bf B}\!\cdot\!\hat{\bf n}$.  This is why a pseudoscalar $F_5$ will
work (and a vector ${\bf F}_\gamma$ will not).  Alternatively, one can
use a pseudovector ${\bf F}_\Sigma$ (e.g., staggered field of an
antiferromagnetic order parameter) to construct a pseudoscalar $({\bf
F}_\Sigma\!\cdot\!\hat{\bf n})$.  

To corroborate these general considerations, I discuss in some detail
several symmetry-breaking terms that might exist in a physical system.

\subsection{Peierls distortion}

We need to add to the $3+1$ dimensional Dirac Hamiltonian
(\ref{Dirac3+1}) a term that breaks $P$ but not $T$ and does not
affect electron spin.  Such a term is
\begin{equation}
V = \tau_2({\bf q}\!\cdot\!\hat{\bf p})
\end{equation}
(both $\tau_2$ and $\hat{\bf p}$ are $T$-odd).  The vector ${\bf q}$
characterizes the direction and length of the Peierls distortion.  It
couples to the operator $-i\vec\gamma\!\cdot\!\hat{\bf n}$
(\ref{gamma-n}).  With the aid of Eq. \ref{15terms},
\begin{equation}
\Delta E 
= \frac{1}{4}\mbox{Tr}
\left(\tau_2({\bf q}\!\cdot\!\hat{\bf p})
(-i\vec\gamma\!\cdot\!\hat{\bf n})\right)
= ({\bf q}\!\cdot\!\hat{\bf p})(\hat{\bf p}\!\cdot\!\hat{\bf n}).
\label{Peierls}
\end{equation}

The charge density is {\em unchanged} when magnetic field is reversed
${\bf B \to -B}$.  By extension (\ref{CS-1CM}), the Hall conductivity is
sensitive to the direction of ${\bf B}$.  For a wall perpendicular to
${\bf q}$, summation over the four L points gives
\begin{equation}
\sigma = \frac{2e^2}{h}\,\mbox{sgn}(eB).
\end{equation}
This describes a perfectly normal Hall current, not a Chern-Simons
current (\ref{CS-1}).  No parity anomaly here.

\subsection{Zeeman effect: uniform magnetic field}

Consider Zeeman interaction $V = -ge ({\bf B}\!\cdot\!\vec\sigma)/2mc$
induced by uniform magnetic field ${\bf B}$ perpendicular to the
domain wall.
\begin{eqnarray}
\Delta E &=& -\frac{ge}{2mc}\,
\frac{1}{4}\mbox{Tr}
\left(({\bf B}\!\cdot\!\vec\sigma)(\vec\Sigma\!\cdot\!\hat{\bf n})\right)
\mbox{sgn}(eB)
= -\frac{g|eB|}{2mc}(\hat{\bf p}\!\cdot\!\hat{\bf n})^2.
\label{Zeeman-uniform}
\end{eqnarray}
At a constant chemical potential, 
\begin{equation}
\frac{Q}{A} = \frac{4e|eB|}{4\pi c}\, \mbox{sgn}(g).
\end{equation}
Again, a normal Hall effect results:
\begin{equation}
\sigma = - \frac{2e^2}{h}\,\mbox{sgn}(geB).
\end{equation}

\subsection{Zeeman effect: staggered magnetic field}

In a system with antiferromagnetic order, one expects staggered
magnetic field ${\bf B_{\rm st}}$ to induce a Zeeman term $V =
-\tau_3\,ge ({\bf B_{\rm st}}\!\cdot\!\vec\sigma)/2mc$.  It couples to
the same operator of Dirac ``spin'' (\ref{Sigma-n}):
\begin{equation}
\Delta E = -\frac{ge}{2mc}\,
\frac{1}{4}\mbox{Tr}
\left(\tau_3 ({\bf B_{\rm st}}\!\cdot\!\vec\sigma)
	(\vec\Sigma\!\cdot\!\hat{\bf n})\right)\mbox{sgn}(eB)
= -\frac{g|e|\mbox{sgn}(B)}{2mc}
([\hat{\bf p}\!\times\![{\bf B}_{\rm st}\!\times\!\hat{\bf p}]]
	\!\cdot\!\hat{\bf n}).
\label{Zeeman-staggered}
\end{equation}
In the simple case with ${\bf B}_{\rm st} = B_{\rm st} \hat{\bf n}$,
\begin{equation}
\frac{Q}{A} = \frac{4e^2B}{4\pi c}\, \mbox{sgn}(g e B_{\rm st})
\end{equation}
and 
\begin{equation}
\sigma = -\frac{2e^2}{h}\, \mbox{sgn}(g e B_{\rm st}).
\end{equation}
The Hall current depends on the direction of the staggered field ${\bf
B}_{\rm st}$ and not of the uniform field ${\bf B}$.  One can call this
a parity anomaly, but clearly there is no magic involved: this is
a Hall effect caused by the staggered magnetic field.  

\subsection{A pseudoscalar?}

Finally, one can imagine coupling directly to the third conserved
quantity (\ref{gamma-05}).  The only perturbation that will do is
\begin{equation}
V = \delta\,\tau_2 (\vec\sigma\!\cdot\!\hat{\bf p}).
\end{equation}
The sign of induced charge depends on the direction of magnetic field:
\begin{equation}
\frac{Q}{A} = - \frac{4e^2 B}{4\pi c}\, \mbox{sgn}(\delta). 
\end{equation}
Hall conductivity is insensitive to the sign of $B$:
\begin{equation}
\sigma = \frac{2e^2}{h}\, \mbox{sgn}(\delta). 
\end{equation}

This term definitely generates a Chern-Simons current.  I must admit
though that I do not understand what perturbation could cause it.  

\section{Concluding remarks}

I have presented a critique of the parity anomaly (Eqs. \ref{CS-0} and
\ref{CS-1}) suggested for $2+1$ dimensional fermions in a narrow-gap
semiconductor.\cite{FDB}  Both symmetry considerations and a correct
continuum treatment of the model indicate that a parity anomaly
requires a pseudoscalar ($P$ and $T$-odd) or pseudovector ($P$-even
and $T$-odd) symmetry-breaking term.  Therefore, a Peierls distortion,
which is a vector ($P$-odd and $T$-even), cannot generate a
parity anomaly, contrary to the suggestion of FDB.

Antiferromagnetic order, which violates the time-reversal symmetry,
could lead to an observable Chern-Simons current in the system.
Unfortunately, the prospects of observing such an effect are rather
dim: not only would this require antiferromagnetism, but also the
N{\'e}el order parameter must be (anti)parallel to the applied magnetic
field.  In a Heisenberg antiferromagnet, ${\bf B}_{\rm st}$ stays
orthogonal to ${\bf B}$, in which case the entire effect vanishes.

In hindsight, this foray into the dreamworld of field theory is a
reminder to a condensed-matter physicist: trust your intuition.
Anomalous Hall current of Dirac fermions, whether directly in $d=2+1$
dimensions or on a domain wall in $d=3+1$, is always caused by
magnetic order present in the system.\cite{Semenoff,Haldane,Nagaosa}

\section*{Acknowledgment.} 

I thank N. Nagaosa, L. P. Pryadko and F. Wilczek for helpful
discussions and B. Jank\'o for hospitality during my stay at Argonne.
Research was supported by DOE Grant No. DE-FG02-90ER40542 and 
DOE Office of Science under Contract No. W-31-109-ENG-38.

\appendix

\section{Dirac fermions from a tight-binding model}
\label{appendix}

Lead and a chalcogen have 2 and 4 electrons in their 3 external $p$
orbitals ($s$ orbitals give deeply lying bands, which can be ignored).
The $p$ bands are thus half filled.  The one-particle Hamiltonian
including lattice potential and spin-orbit interaction is
\begin{equation}
H = -\frac{\nabla^2}{2m} + V({\bf r}) 
+ \frac{[\vec\sigma\!\times\!\nabla V({\bf r})]\cdot(-i\nabla)}{4m^2c^2}.
\label{H-full}
\end{equation}
Omitting spin for clarity, 
one can write an energy eigenfunction in the Bloch form
\begin{equation}
\Psi_{\bf p}({\bf R}) 
= \sum_{i}^{p_x,p_y,p_z}\ \sum_{\alpha}^{\rm Pb,Te}\ 
\psi^{\alpha i}({\bf R-r-\Delta r}^\alpha)\, 
\phi^{\alpha i}_{\bf p}({\bf r})
e^{i{\bf p}\cdot({\bf r + \Delta r}^\alpha)}.
\label{Psi}
\end{equation}
Here ${\bf r}$ labels unit cells in the rocksalt lattice, $\Delta {\bf
r}^\alpha$ are the coordinates of site $\alpha$ within a unit cell and
$\psi^{\alpha i}({\bf R})$ is the wavefunction of the $i$-th orbital
on sublattice $\alpha$.  We are to find the Bloch coefficients
$\phi^{\alpha i}_{\bf p}({\bf r})$.  With spin included, $\phi^{\alpha
i \sigma}$ is a 12-component wave function.

It follows from symmetry considerations alone\cite{Mitchell-Wallis}
that the two sublattices (lead and chalcogen sites) are decoupled at
the L points, such as ${\bf p} = \pm(\pi/a)(1,1,1)$.  In this case,
the Hamiltonian takes on a simple form\cite{VPS}
\begin{equation}
H_\alpha = 2W_\alpha \cos{\left(\frac23{\bf L\!\cdot\!\hat{p}}\right)}
+ \lambda_{\alpha}\vec\sigma\!\cdot\!{\bf L},
\label{H0}
\end{equation}
where the first term comes from the overlap of different $p$ orbitals
(between second neighbors) and the second, from the spin-orbit
interaction.  The eigenstates of the Hamiltonian (\ref{H0}) are
Kramers doublets with ${\bf J\!\cdot\!\hat{p}} = \pm 1/2$ or $\pm
3/2$.  According to Volkov {\em et al.},\cite{VPS} the spin-orbit
coupling is the weakest perturbation and the two bands around the
Fermi level are derived from states with ${\bf L\!\cdot\!\hat{p}}=0$.
In this approximation,\cite{non-essential} the eigenstates are
\begin{equation}
\phi^{\alpha \sigma}_{\bf p}({\bf r}) 
= \frac{1}{\sqrt{3}}\left(
\begin{array}{r}
1 \\ 1 \\ 1
\end{array}
\right) \otimes |\sigma_z = \sigma\rangle, 
\label{jz-states}
\end{equation}
in the basis of $p_x, p_y, p_z$ orbitals.

To calculate the matrix elements of the Hamiltonian near an L point,
at lattice momentum ${\bf p+k}$, I use the standard ``${\bf
k\!\cdot\!p}$'' method.\cite{Mitchell-Wallis} Instead of shifting
momentum away from a symmetry point ${\bf p \to p+k}$ in the
wavefunctions (\ref{Psi}), one transforms the Hamiltonian
\begin{equation}
H \to e^{-i{\bf k\cdot R}} H e^{i{\bf k\cdot R}}
= H + \frac{1}{m}\, {\bf k}\!\cdot\!\left(
-i\nabla + \frac{\vec\sigma\!\times\!\nabla V({\bf R})}{4mc^2}
\right).
\end{equation}
The matrix elements of the second term ``$({\bf k\!\cdot\!p})/m$'' are
then evaluated in the Hilbert space of the two Kramers doublets 
closest to the Fermi level at the L
point.  This is a good approximation for small ${\bf k}$, provided
that other states lie at energies much larger than the gap
$\Delta = (E_{\rm Pb} - E_{\rm Te})/2$.

\begin{figure}
\centerline{
\epsfxsize 3in
\epsffile{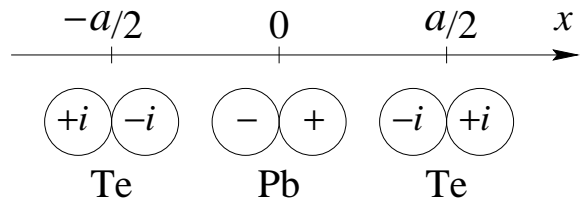}
}
\caption{Calculation of overlap integrals for adjacent $p_x$ orbitals.
Momentum $p_x=\pi/a$.}
\label{fig-px}
\end{figure}

In the tight-binding limit, overlap of $p_x$ orbitals on different
sublattices occurs along the $x$ direction only.  In the case of $p_x$
orbitals at momentum $p_x=\pi/a$ (Fig. \ref{fig-px}), operators
$-i\nabla$ and $\nabla V({\bf R})$ have the following nonzero matrix
elements:
\begin{eqnarray}
\langle \mbox{Pb}, x=0 | 
(-i\partial_x) | \mbox{Te}, x=a/2\rangle &=& C_1,\\ 
\langle \mbox{Pb}, x=0 | 
\partial_x V | \mbox{Te}, x=a/2\rangle &=& iC_2, 
\end{eqnarray}
where $C_1$ and $C_2$ are some real constants.  In the basis of states
$p_x$, $p_y$, $p_z$, the resulting ${\bf k\!\cdot\!p}$ term reads
\begin{equation}
C_1\tau_1\left( 
\begin{array}{ccc}
k_x & 0 & 0 \\
0 & k_y & 0 \\
0 & 0 & k_z
\end{array}
\right)
+ C_2\tau_2\left( 
\begin{array}{ccc}
k_y\sigma_z-k_z\sigma_y & 0 & 0 \\
0 & k_z\sigma_x-k_x\sigma_z & 0 \\
0 & 0 & k_x\sigma_y-k_y\sigma_x
\end{array}
\right)
\end{equation}
near the L point ${\bf p} = (\pi/a)(1,1,1)$.  The operators of isospin
$\tau$ act according to
\begin{equation}
\begin{array}{cccc}
\tau_1 |\mbox{Pb}\rangle = |\mbox{Te}\rangle, & 
\tau_2 |\mbox{Pb}\rangle = i|\mbox{Te}\rangle, & 
\tau_3 |\mbox{Pb}\rangle = |\mbox{Pb}\rangle, & 
\tau_3 |\mbox{Te}\rangle = -|\mbox{Te}\rangle.
\end{array}
\end{equation}

Matrix elements of the complete Hamiltonian between the states
(\ref{jz-states}) are:
\begin{eqnarray}
& \frac{1}{3}C_1\tau_1(k_x+k_y+k_z)
+ \frac{1}{3}C_2\tau_2\left[
(k_y-k_z)\sigma_x
+ (k_z-k_x)\sigma_y
+ (k_x-k_y)\sigma_z
\right] &
\nonumber \\
&+\frac{1}{2}(E_{\rm Pb} - E_{\rm Te}) \tau_3.&
\end{eqnarray}
This yields the anticipated continuum result (\ref{Dirac3+1})
\begin{equation}
H = v ({\bf k\!\cdot\!\hat{p}})\, \tau_1 
+ \lambda v ({\bf k\!\cdot\![\hat{p}\!\times\!\vec\sigma]})\, \tau_2 
+ Mv^2\, \tau_3,
\end{equation}
where $\hat{\bf p}$ is the unit vector in the direction of the L 
point ${\bf p} = (\pi/a)(1,1,1)$.  An identical result 
is obtained for the 3 other L points, which differ by the direction
$\hat{\bf p}$. 

On a final note, the unit vector $\hat{\bf p} = {\bf p}/p$ is odd
under parity and time reversal defined in Eq. \ref{PT}.  It can 
be made even by redefining them as
\begin{equation}
\begin{array}{cc}
P \psi({\bf r}) = -\tau_3\psi(-{\bf r}), &
T \psi({\bf r}) = \tau_3\sigma_y \psi^*({\bf r}).
\end{array}
\label{PT-conventional}
\end{equation}
This is the familiar standard representation of parity and time reversal
in relativistic field theory.\cite{Landau}

\begin{thebibliography}{99}
\bibitem{non-essential} The form of the Hamiltonian in the continuum
limit (\ref{Dirac3+1}) is strongly constrained by symmetry
requirements.  It is therefore not sensitive to the approximation made
here.  I have checked that the same result holds in the opposite
limit, when the spin-orbit coupling in Eq. \ref{H0} dominates.  
Then $\vec{\sigma}$ is the total angular momentum of an electron.
\end{thebibliography}
\end{document}